\documentclass[prb,twocolumn,showpacs,showkeys]{revtex4}
\usepackage{graphicx}
\usepackage{graphics}
\usepackage{dcolumn}
\usepackage{bm}
\newcommand{\beq}{\begin{equation}}
\newcommand{\eeq}{\end{equation}}
\newcommand{\beqnar}{\begin{eqnarray}}
\newcommand{\eeqnar}{\end{eqnarray}}
\newcommand{\bfig}{\begin{figure}[!hbp]}
\newcommand{\efig}{\end{figure}}
\begin{document}

\title{Scaling properties of one-dimensional off-diagonal disorder}

\author{Hosein Cheraghchi$^{1,2}$} 

\affiliation{$^{1}$Department of Physics, Sharif University of
Technology,P.O.Box 11365-9161, Tehran, Iran \\ $^{2}$Department of
Physics, Damghan University of Basic Sciences, Damghan, Iran}
\email{cheraghchi@mehr.sharif.edu}
\date{\today}

\begin{abstract}
Validity of the single parameter scaling (SPS) in one dimensional
Anderson model with purely off-diagonal disorder is being studied.
It is shown that the localized region with standard symmetry is
divided into two regimes: SPS and non-SPS. Scaling relations of
the Lyapunov Exponent are proposed for these two regimes. In the
non-SPS regime, in additional to the localization length, there
exists a new length scale which is related to the integrated
density of states. A physical interpretation of the new length is
the cross-over length which separates regions with chiral symmetry
from those that have standard symmetry.

\end{abstract}


 \maketitle
 \section{Introduction}
It is well-known for about forty years that all electron states in
standard one dimensional (1D) disordered models are localized for
any strength of disorder, and there is no localization transition
in 1D systems [\onlinecite{anderson}]. However, in the case of
off-diagonal disorder, there is an anomalous localized state at
the band center [\onlinecite{soukoulis,cohen,inui,ziman}]. It has
been proposed that the Lyapunov Exponent (L.E.) is the appropriate
scaling variable to describe fluctuations of the conductivity.

\beq\gamma(N) = \frac{1}{2N}\ln(1+\frac{1}{g}) = -
\frac{1}{2N}\ln(T) \label{eq:gama} \eeq
where $g (= T / R)$ and $T$ are conductance and transmission coefficients through the system
with length $N$.

A reason for the revival of the interest in 1D disordered model
is due to the revision of the well-known Single Parameter Scaling
(SPS) hypothesis. According to this hypothesis
[\onlinecite{abrahams}], there exists a single parameter,
conductance $g$, which determines scaling properties of $g(N)$.
Soon after the report of SPS, it became clear that one should
consider scaling of the full probability distribution of
conductivity.

In order to take fluctuations of conductance into account, one
should consider a parameter $\gamma$ (L.E.) defined in
Eq.(\ref{eq:gama}) instead of conductance $g$ itself. In the
thermodynamic limit, $\gamma_{0} = \gamma({N\rightarrow\infty})$
has a non-random value which is the inverse of the localization
length $\lambda$. This parameter has normal distribution for $ N
\gg \lambda$, and its dispersion $\sigma_{\gamma}$ obeys the law
of large numbers and is not an independent variable.

 \beq \frac{1}{\tau} =
\frac{\gamma_{0}}{N \sigma^{2}_{\gamma}}=1
\label{eq:measure}\eeq
where the parameter $\tau$ is usually defined as a conventional scaling
parameter in literature. The above equation was originally
derived by Anderson {\it et al}.[\onlinecite{Andersonscale}] by using
the random phase hypothesis.

However, as shown in [\onlinecite{Deych}] without the assumption
of phase randomization, 1D SPS (Eq.[\ref{eq:measure}]) is
violated, where states are much far apart from each other than the
localization length ($\lambda$).

 This new characteristic length scale $\ell_{s}$ which is related to the distance between states,
for the states near the band center is defined in terms of total
number of states ($N(E)$) whose energy is less than $E$
[\onlinecite{Deych2}]. This criterion was initially extracted from
the exact calculation of the variance of L.E. for the Anderson
model with Cauchy distribution of the site energies
[\onlinecite{Deych}]. However in contradiction with violation of
SPS in 1D [\onlinecite{schomerus}], it has been recently shown
using exact diagonalization [\onlinecite{kantelhardt}] and
transfer matrix method [\onlinecite{Queiroz}] that at special
point $E=0$, SPS holds perfectly in 1D.

In 2D case, the interest is motivated by the experimental
observations of a metal-insulator transition which is at odds with
the SPS for noninteracting electrons [\onlinecite{kravchenko}].
The validity of SPS in 2D is currently very controversial. There
exists some numerical analysis of 2D Anderson model which confirms
the SPS hypothesis [\onlinecite{Kramer, Scheriber,slevin}]. Other
studies suggest a two-parameter scaling
[\onlinecite{kantelhardt,Queiroz,Prior}]. It has been shown
[\onlinecite{Prior}] that 2D SPS does not follow
Eq.(\ref{eq:measure}).

In 1D systems with off-diagonal disorder (random hopping model),
it is clear that an anomalously localized state at $E=0$, results
in a violation of SPS. Divergence of the localization length and
density of states at the band center in this model
[\onlinecite{Brouwer1},\onlinecite{Brouwer3}], is in contradiction
with the scaling theory. Unusual properties of this model are due
to chiral symmetry [\onlinecite{Brouwer2,Mudry}]. In an interval
close to an anomalous state, SPS does not hold
[\onlinecite{Deych2}]. The main objective of the present paper, is
to answer how far from the anomalous state (at $E=0$), 1D SPS will
again be held.

In this paper, with care of some debates on the validity of 1D
SPS, we reexamine the scaling properties of one-dimensional system
with purely off-diagonal disorder by using transfer matrix method.
Our attention is on a region near the band center which contains
strongly localized states with standard symmetry. In the strong
localization limit, it will be shown that the L.E. distribution
function is normal. In this region, it is shown that there exists
a new length scale which is the same as the length scale
($\ell_{s}$) defined in Eq.(\ref{eq:ells}) [\onlinecite{Deych}].
The SPS exists as long as the localization length $\lambda$
exceeds $\ell_{s}$. In the SPS region, it is shown that the L.E.
only depends on the disorder strength as
$\gamma_0\propto\sigma_{\ln(t)}^2$. The scaling properties of the
non-SPS region which has been reported in Ref.
[\onlinecite{cheraghchi}], is confirmed by a data collapse. The
variance and mean of the L.E. for different disorder strengths,
system sizes and also for various range of energy spectrum, lie on
a single curve when they are expressed in terms of the scaling
parameter ($\tau$) defined in Eq.(\ref{eq:measure}) as a function
of the ratio $\kappa=\lambda/\ell_{s}$. It can also provide a
physical interpretation for $\ell_{s}$ as a cross-over length
between two chiral and standard symmetries.

 This article is organized as follows: Section {\bf II} describes our model and the
exact calculation of all L.E. moments at the band center. Section
{\bf III} describes the transition from chiral symmetry region to
localized region by the calculation of L.E. distribution function
and its mean. In this Section, it will be shown that the localized
region is divided into non-SPS and SPS regimes. A new length
scale($\ell_{s}$) which controls the scaling theory is defined in
Section (IV). We try to find a meaningful physical interpretation
for the new length scale as a cross-over length in Section ({\bf
V}). Discussions and conclusions are finally presented in Section
{\bf VI}.

\section{Model and Moments of Lyapunov Exponent}
We consider non-interacting electrons in 1D disordered systems
within a tight binding approximation. The Schroedinger equation
with the assumption of nearest-neighbor hopping becomes
 \beq
\varepsilon_{i}\psi_{i}+t_{i,i+1}\psi_{i+1}+t_{i-1,i}\psi_{i-1}=E\psi_{i}
\eeq where E is the energy corresponding to the electron wave
function. ${\bf |\psi_{i}|}^{2}$ is the probability of finding the
electron at site i, ${\varepsilon_{i}}$ are the site potentials
and ${t_{i-1,i}=t_{i,i-1}=t_{i}}$ the hopping terms. Using the
transfer matrix method, one can relate the electron wave functions
at the two ends of the system to each other. In our model, we
consider all site energies to be zero and a periodic boundary
condition on hopping terms as $t_{1} = t_{N+1}$. All energies
which appear, are scaled by typical mean of hoppings terms $t_0$,
where $\ln(t_{0}) = <\ln(t_{i})>_{c.a.}$. Here, {\bf c.a.} refers
to the configurational average. The L.E. can be extracted from the
eigenvalues of the total transfer matrix
[\onlinecite{cheraghchi}].

As proved in Ref.[\onlinecite{cheraghchi}], the L.E.
 at $E=0$ has a {\it semi-Gaussian} distribution with a mean which
 can be derived in terms of the pair correlation function. By having the distribution function, higher powers of the L.E. can
 be simply derived in the case of correlated and uncorrelated disorder at the band center ($E=0$) as:

 \begin{eqnarray} <\gamma^2>=\frac{\pi}{2}<\gamma>^{2} ;
<\gamma^3>=\pi <\gamma>^{3} ;   ... ; \nonumber\\
<\gamma^n>\propto<\gamma>^{n}
\end{eqnarray}
Therefore, higher moments of the L.E. can be written as:

\begin{eqnarray}\sigma^{2}_{\gamma}=<(\gamma-<\gamma>)^{2}>=(\frac{\pi}{2}-1)<\gamma>^{2}
\nonumber \\
<(\gamma-<\gamma>)^{3}>=(2-\frac{\pi}{2})<\gamma>^{3}
\label{eq:moments}\end{eqnarray}
%
%
\bfig
\includegraphics[width=8cm]{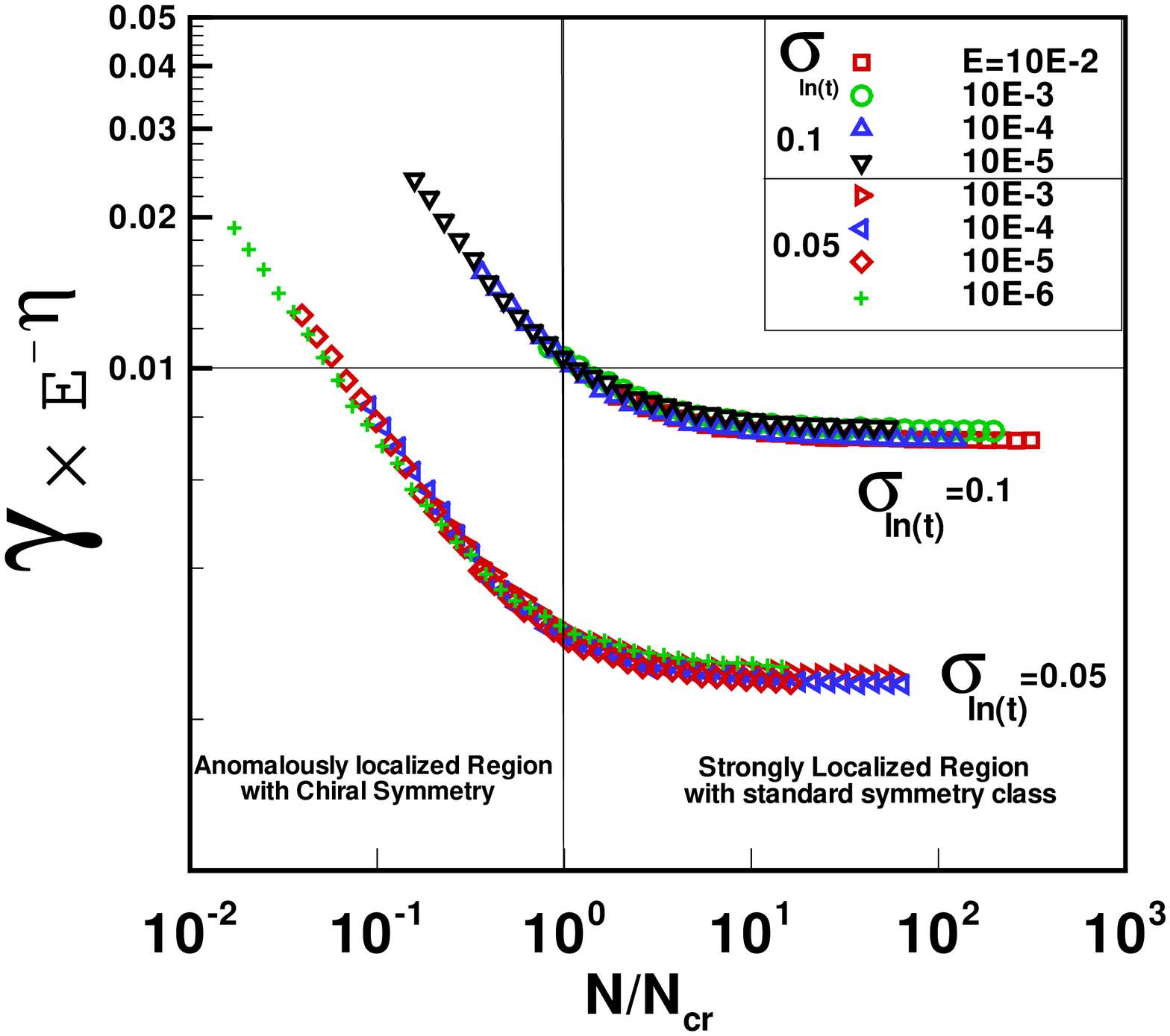}
\caption{ The size dependency of the Lyapunov Exponent for
different energies with $\eta=0.18$.\label{fig:gamma(N)}}
 \efig
This can be generalized to the n'th moment of the L.E. which will
be as $<(\gamma-<\gamma>)^{n}>\propto<\gamma>^{n}$. By paying our
attention to the L.E. form at $E=0$ ($<\gamma>
\propto\sigma_{\ln(t)}/N^{1/2}$), it can be seen that the variance
of the L.E., scales according to the law of large numbers for
uncorrelated disorder. As it has been mentioned by Anderson. et al
[\onlinecite{Andersonscale}], the localization properties can be
described by a variable (such as L.E.) whose width of distribution
function follows the law of large numbers.
 \beq
\sigma_{\gamma}^{2}=(1-\frac{2}{\pi})\frac{\sigma_{\ln(t)}^{2}}{N}\label{eq:unvar}\eeq
This equation and its equivalence in Eq.[\ref{eq:moments}] are
consistent with the result of Ref.[\onlinecite{Mudry}], where it
was derived for a weak disorder by solving the Fokker-Planck
equation. However, the size dependence of the L.E. variance will
change when the disorder becomes correlated.
 As a result, the L.E. distribution function and all
its higher moments converge for large system sizes. So, L.E. is a
good variable to describe statistical properties of disordered
systems.

\section{Scaling and distribution function of Lyapunov Exponent}
\subsection{\label{sec:level31} Scaling of Lyapunov Exponent}
 We calculate the mean and variance of L.E. by using
 the transfer matrix method when randomness is imposed on
 $\ln(t)$'s. The study of L.E. close to the band center results in the coexistence
 of two symmetries in this system. It can be shown that there is a chiral symmetry at
 $E=0$. This is a significant property of purely
off-diagonal disorder with nearest-neighbor approximation.
However, at energies close to the band center, and for lengths
greater than a cross-over length ($N_{cr}$), chiral symmetry is
broken. At sufficiently long lengths, localization properties will
flow to those of the standard symmetry class. All states in this
regime are strongly localized.

For any realization of the disorder, the energy density of states
is symmetric around the band center. This symmetry, which
originates from the fact that the disorder preserves the
bipartite structure of the lattice, is referred to as chiral
symmetry. The chiral symmetry is broken by, e.g., on-site
randomness or next-nearest-neighbor hopping
[\onlinecite{inui,Mudry}].

In the case of purely onsite disorder case which was originally
considered by Anderson [\onlinecite{anderson}], one distinguishes
three universality classes, corresponding to the presence or
absence of time reversal and spin-rotation symmetry. These three
classes are called orthogonal, unitary, and symplectic
[\onlinecite{Mudry}]. Here, we will refer to these as the three
standard universality classes. In this paper, it will be shown
that the standard symmetry region (strongly localized region) is
also divided into two regimes (non-SPS and SPS regimes defined in
section (III.C) and Fig.(\ref{fig:Nvar-gam})), which depend on the
number of scaling parameters.

Fig.(\ref{fig:gamma(N)}) shows the scaling properties of L.E. near
the band center. All data with various energies lie on a single
curve when ($\gamma\times E^{-\eta}$) is plotted in terms of the
dimensionless variable ($N/N_{cr}$). It confirms a power law
divergence of the localization length where energies belong to the
non-SPS regime (Fig.(\ref{fig:Nvar-gam})). Therefore, the
following scaling law of L.E. at $E \neq 0$ can be proposed
[\onlinecite{cheraghchi}].

 \beq (\gamma\times E^{-\eta}) \propto \left\{
\begin{array}{c} \sigma_{ln(t)}^2(\frac{N}{N_{cr}})^{-1/2}
\,\,\,\,\,\,\,\,\,\,\,\,\,\,\,\,\,\ N /N_{cr}\ll1
\\ \\ \sigma_{ln(t)}^{2}
\,\,\,\,\,\,\,\,\,\,\,\,\,\,\,\, N/N_{cr}\gg1\end{array} \right.
\label{eq:cr-energy}
 \eeq
 where the cross-over length is as
 \beq N_{cr}\propto{E^{-2\eta}}/{\sigma_{ln(t)}^{2}}\eeq
  and $\eta \approx 0.18 \pm 0.03$. However, for energies
 very close to the band center, $\eta$ has a small energy dependence. Fig.(\ref{fig:gamma(N)})
 shows a transition from the region with chiral symmetry ($N \ll N_{cr}$) to the region with standard
symmetry ($N \gg N_{cr}$). It was also checked that each of the
data sets do not collapse on each other when one uses the
logarithmic energy dependence of the localization length as seen
in Refs.(\onlinecite{cohen,ziman}).

\subsection{\label{sec:level32} Distribution Function of Lyapunov Exponent}

The localization properties of different symmetry regions can be
also characterized by the distribution function of L.E. As it was
mentioned in section (II), at the band center, the distribution
of L.E. is semi-Gaussian. For zero energy ($E=0$), numerical
evidence in Fig.(\ref{fig:distribution}.a) confirms such
analytical distribution function for all system sizes.
%
%

 \bfig
 \begin{center}
\includegraphics[width=8 cm]{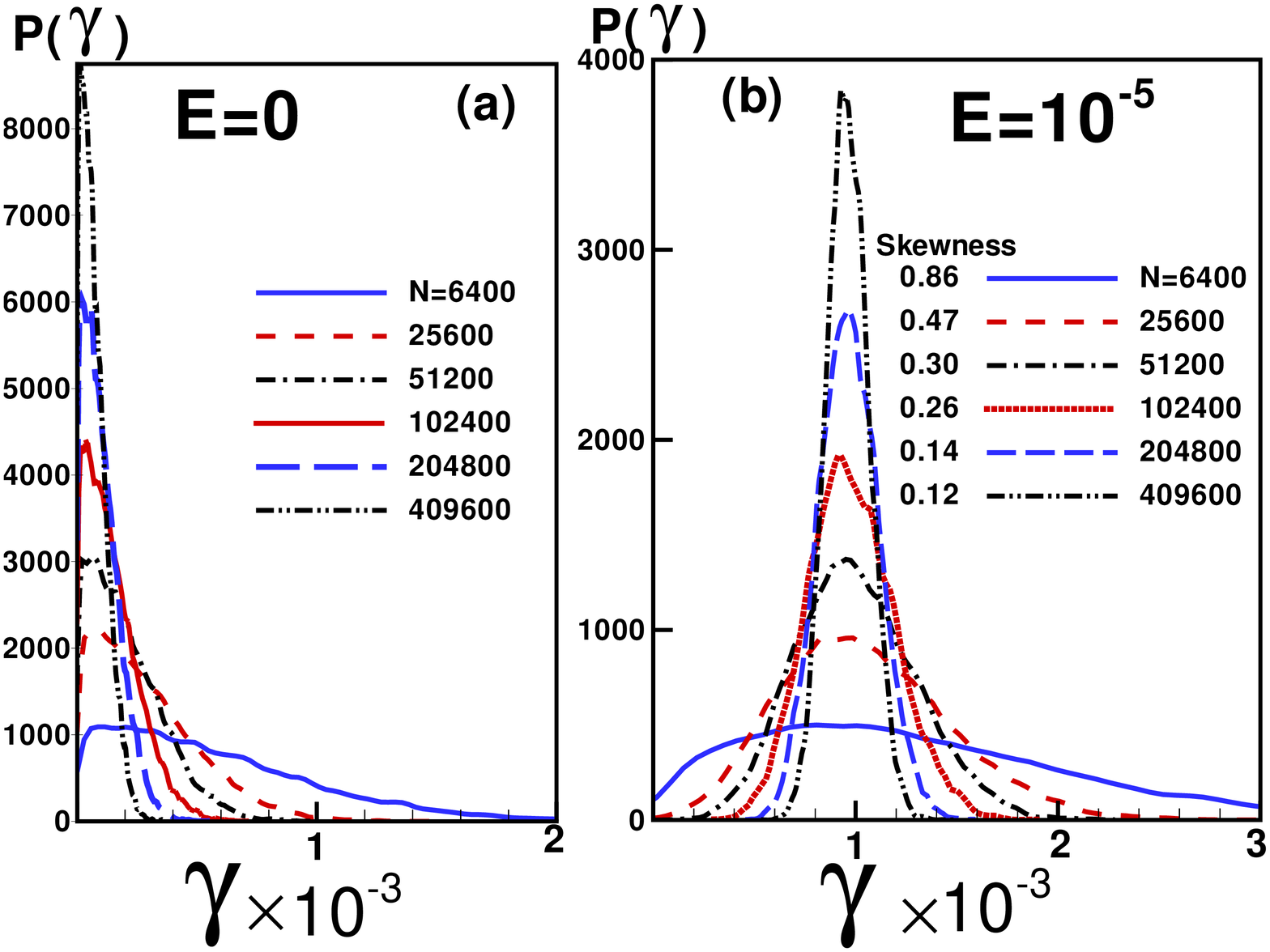}
\end{center}
\caption{Lyapunov Exponent Distribution Function for different
system sizes at a) $E=0$ and b) $E=10^{-5}$. Disorder strength is
considered to be 0.1. The number of samples is $2\times10^4$
configurations. \label{fig:distribution}}
 \efig

 For energies near the band center ($E\neq0$), and for system sizes smaller
than the cross-over length ($N \ll N_{cr}$), the distribution of
L.E. is {\it semi-Gaussian}. In this region, the band center
behavior is dominant. However, as the system size increases, the
distribution
function becomes more \textit{Gaussian}-like for $N \gg N_{cr}$. 
This region has a regular Anderson-like behavior with standard
symmetry class. Fig.(\ref{fig:distribution}b) shows distribution
function of the L.E. for the same system sizes as the band center
case. It can be seen that there exists again a transition between
these two symmetries; from chiral to standard symmetry or from
semi-Gaussian to Gaussian distribution. For the sake of
completeness, the skewness of the distribution functions has been
calculated as a measure of the symmetry of the distribution. It
can be defined as [\onlinecite{press}]: \beq {\rm
Skewness}=\frac{<(\gamma-{\overline{\gamma}})^3>}{{<(\gamma-{\overline{\gamma}})^2>}^{3/2}}\eeq
 Distribution close to the normal form, has a skewness equal to zero. A distribution whose skewness has absolute
value less than $0.5$ is considered fairly symmetrical. Therefore,
distributions of long enough systems in
Fig.(\ref{fig:distribution}b), are very close to the Gaussian
form. In the SPS regime (Fig.(\ref{fig:Nvar-gam})) where the L.E.
is independent of size and energy, distribution function of L.E.
is exactly Gaussian (the skewness order of $10^{-3}$) and
independent of size. All distribution curves have been softened by
the Kernel smoothing method [\onlinecite{Kernel}] without changing
any statistical characteristic of distributions.
\subsection{\label{sec:level33} SPS and Non-SPS Regimes}
For a fixed length ($N$), Eq.(\ref{eq:cr-energy}) proposes a
critical energy point which separates two regions with different
symmetries. In fact, for energies greater than
$\varepsilon_{cr.}^{(1)}\propto (N
\sigma_{\ln(t)}^{2})^{\frac{-1}{2\eta}}$, the system is in the
localized region. Fig.(\ref{fig:Nvar-gam}) shows the energy
dependence of L.E. in the localized region. It can be seen that
there is a second critical point ( $\varepsilon_{cr.}^{(2)}\approx
10^{-2}$ in Fig.(\ref{fig:Nvar-gam})) where the energy spectrum is
divided into SPS and non-SPS regimes. In the SPS regime, L.E. is
independent of energy and the scaling theory is valid
(Eq.(\ref{eq:measure})). Although, near the band edges anomaly,
L.E. will become energy dependent. In the SPS regime, it has been
checked (Figs.(\ref{fig:gamma(N)},\ref{fig:Nvar-gam})) that in
contradiction
 to the result of Ref.[\onlinecite{izrailev}], the
L.E. only depends on disorder strength. Fig.(\ref{fig:Nvar-SN})
shows that the L.E. is proportional to the square of disorder
strength.
\beq
\gamma_0\propto\sigma_{\ln(t)}^2\label{eq:sps_scale}\eeq
The line Fitted on data in Fig.(\ref{fig:Nvar-SN}) has a slope equal to the value 2.
The coefficient of the above scaling law is ($\frac{1}{3.00\pm0.2}$)
which can be extracted for a
fixed disorder strength ($\sigma_{\ln(t)}=0.1$) and for an energy
($\varepsilon=0.1 \geq\varepsilon_{cr.}^{(2)}$) in the SPS regime.
The skewness of this point is about 0.005. This skewness shows an exactly
Gaussian form for L.E. in this regime.

It can be seen that matching of two
Eqs.(\ref{eq:cr-energy},\ref{eq:sps_scale}) at the boundary of
non-SPS to SPS regime ($E=\varepsilon_{cr}^{(2)}$) leads to a
second critical point ($\varepsilon_{cr}^{(2)}$) which is
independent of all system parameters (a constant).
%
%
\bfig
\includegraphics[width=8 cm]{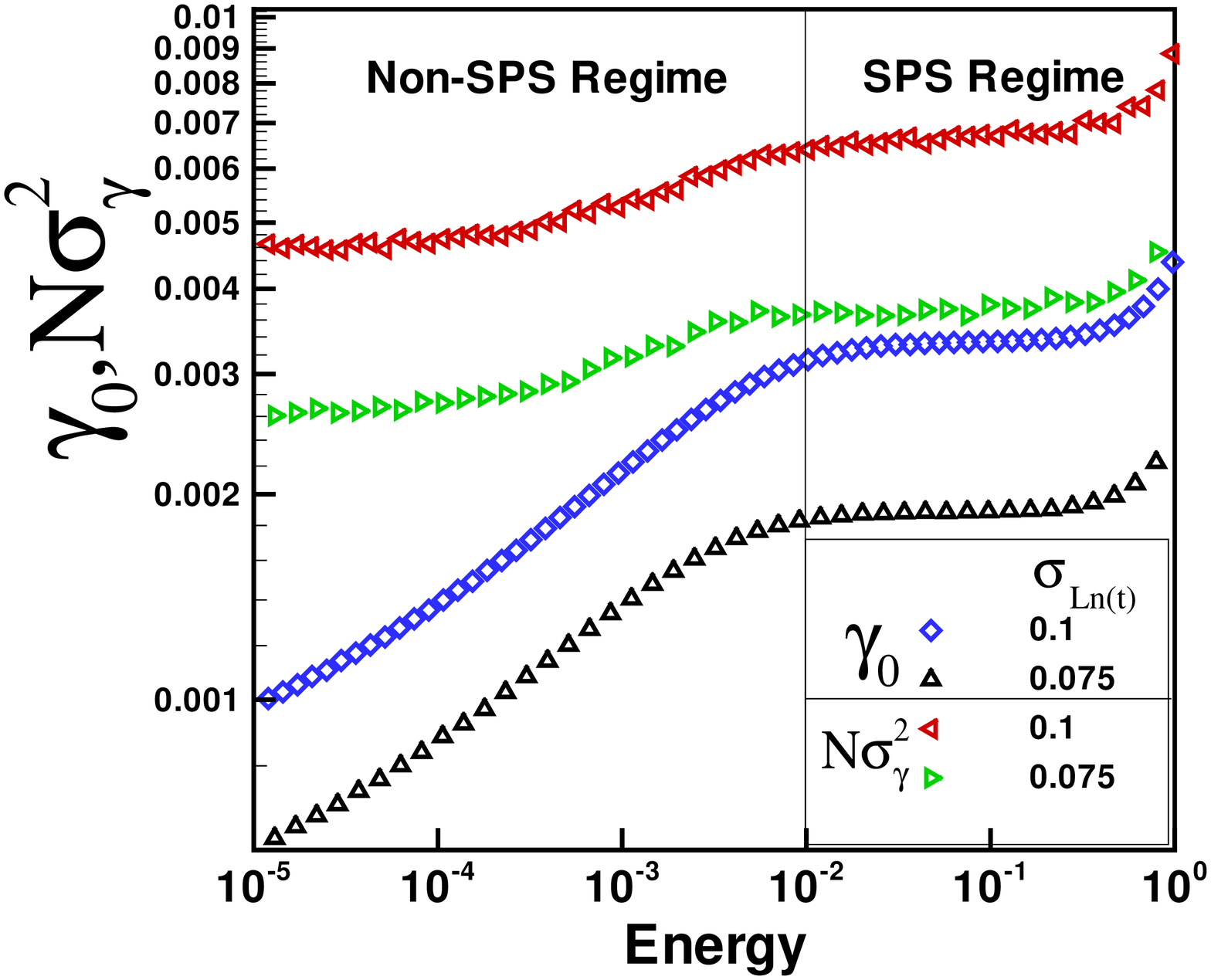}
\caption{SPS to non-SPS transition with plotting Lyapunov Exponent
and $N\sigma_{\gamma}^{2}$ versus energy for system size
$10^5$. $\varepsilon_{cr.}^{(1)}$ which depends on the disorder strength,
is about $5\times10^{-9}$ and $3\times10^{-7}$ for $\sigma_{\ln(t)}=0.1$ and $0.075$
, respectively. $\varepsilon_{cr.}^{(2)}$ is equal to $10^{-2}$.\label{fig:Nvar-gam}}
 \efig

Now, we numerically study the variance of L.E. as a function of
system parameters. Fig(\ref{fig:Nvar-gam}) shows
$N\sigma_{\gamma}^{2}$ versus energy at fixed system size. As it
can be seen, the variance of L.E. is approximately independent of
energy at energies in the SPS and also non-SPS regimes. Figs.(\ref{fig:Nvar-gam},\ref{fig:Nvar-SN}) show that size
and disorder strength dependence of the variance of L.E. follows
from Eq.(\ref{eq:unvar}). As it is clear from a fitted line (with slope $2$)
 on data in Fig.(\ref{fig:Nvar-SN}),
the quantity $N\sigma_{\gamma}^2$ is
proportional to $\sigma_{\ln(t)}^{2}$. The size dependence of the
L.E. variance ($\sigma_{\gamma}^{2}(N)$) is shown in the inset
Fig.(\ref{fig:Nvar-SN}). The L.E. variance decreases with the
inverse of the system size similar to the size dependence of
variance at the band center (Eq.(\ref{eq:unvar})). So, the
variance of L.E. at the band center can be generalized to other
energies near the band center.
\section{Violation of Single Parameter Scaling}
According to Eq.(\ref{eq:measure}), the two parameters of the
distribution reduce to only one. Parameter $\tau$ can be defined
as a measure of SPS in that equation.
%
%
\bfig
\includegraphics[width=8 cm]{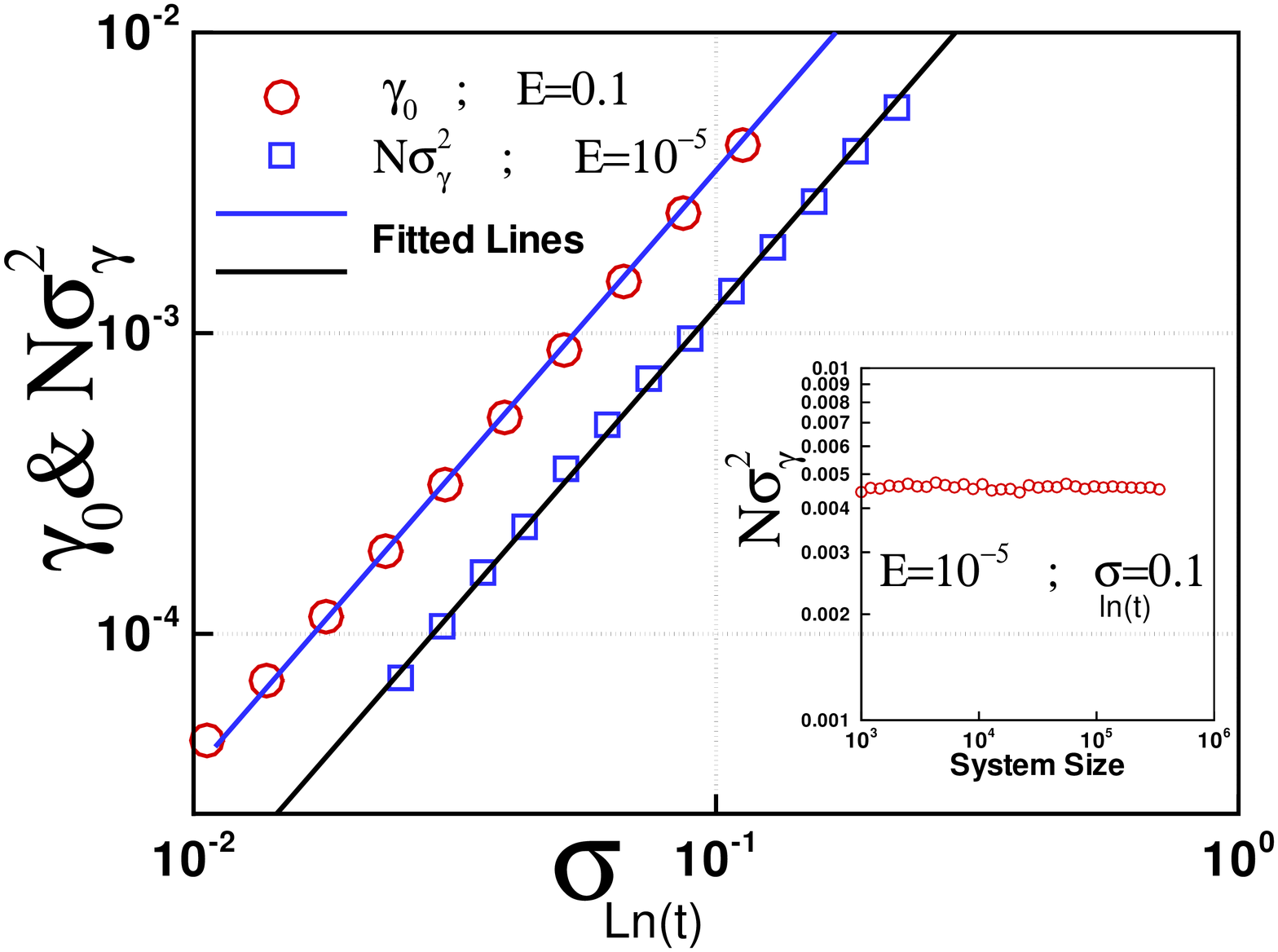}
\caption{ Standard deviation of the Lyapunov Exponent versus
disorder strength for system with length $10^5$. Dependency of
$\gamma_0$ to disorder strength in the SPS regime ($E=0.1$). Inset
figure shows the system size dependence of the variance of
Lyapunov Exponent.\label{fig:Nvar-SN}}
 \efig
Now, in this Section, we try to find the independent parameters of
the system in non-SPS regime. It is again stressed that the random
hopping model away from the $E=0$ and in the localized region is
being studied. The quantity of interest is $\ell_{s}$ that is
related to the integral density of states from the band center to
a given energy, normalized by the total number of states in the
band.

\beq \ell_{s}=\frac{1}{\sin(\pi N(E))} \label{eq:ells}\eeq

For the Anderson model, $N(E)$ can be computed by the node-counting
theorem [\onlinecite{node}]. By starting an initial vector in
transfer matrix method, we count the number of wave function nodes
as the length is scanned.

Fig.(\ref{fig:SPS-D}) shows the numerical result of inverse
scaling parameter ($\tau (\kappa)$) in terms of the dimensionless
parameter ($\kappa=\frac{\lambda}{\ell_{s}}$) for different values
of disorder strength and energy.

 The data included in this graph correspond to the localized regime
 with standard symmetry class where the L.E. has a Gaussian distribution.
 The inset Fig.(\ref{fig:SPS-D}) shows that the skewness of
the data are less than 0.25 for $\kappa>0.1$. All data are in the
region $N\gg\ell_{s}$. What is important, is that all data with
different values of energy and disorder strength and also system
size collapse to a single curve when they are expressed in terms
of $1/\tau$ and $\kappa$. Therefore, for $\kappa \ll1$ (non-SPS
regime) variance of L.E. depends on two parameters; $\kappa$ and
the mean of L.E. In the case of $\kappa \gg1$,
the inverse of the scaling parameter ($1/\tau$) in the present model
goes to the value 0.5 which is different from unity. Therefore, our expression
from non-SPS is only the deviation of Eq.(\ref{eq:measure}) from 0.5.
%
%
\bfig
\includegraphics[width=8 cm]{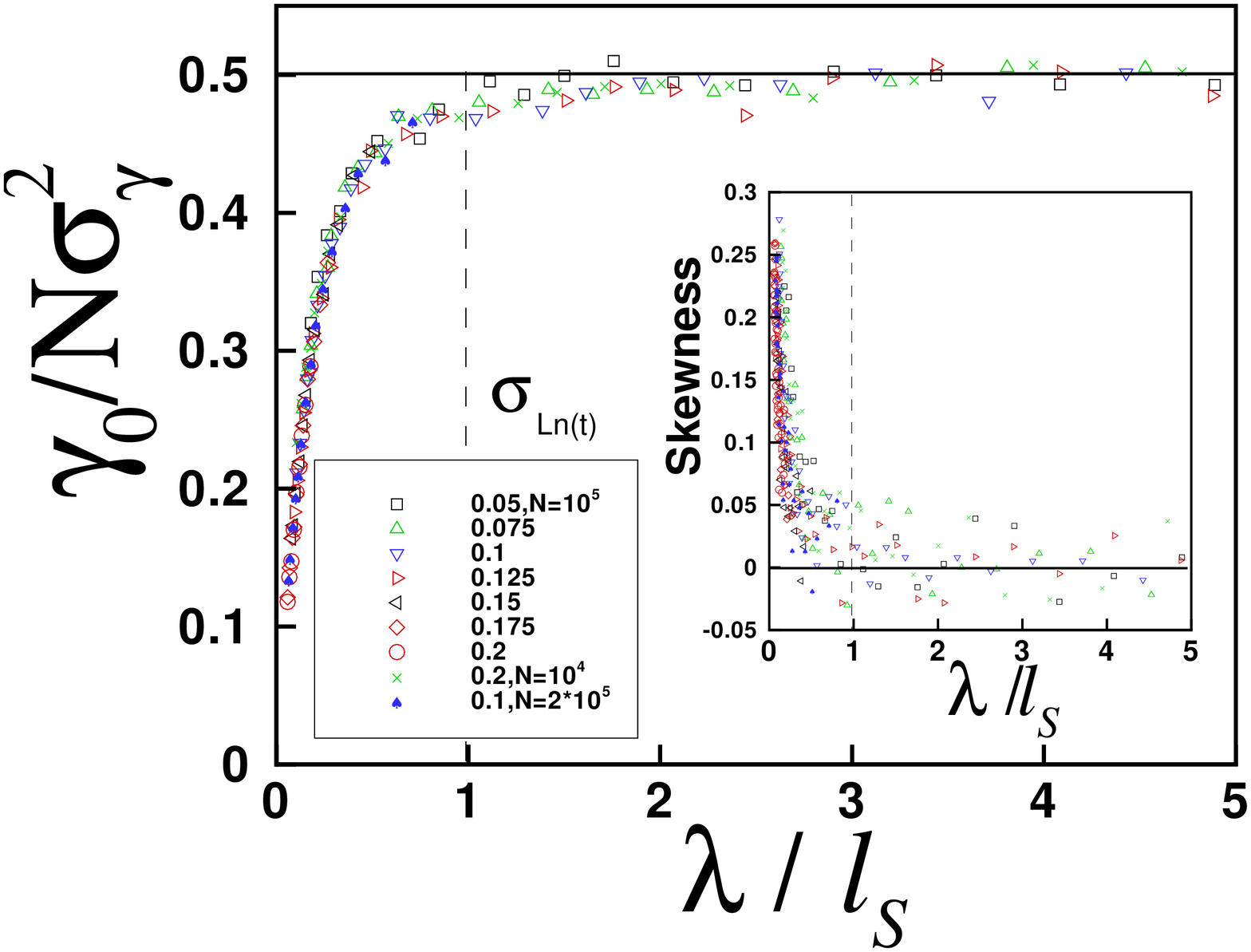}
\caption{ Measure of the validity of the Single Parameter Scaling
by inverse Scaling Parameter ($1/\tau =\gamma_{0}/(N
\sigma^{2}_{\gamma})$) in terms of $\kappa=\lambda/\ell_{s}$ for
disorder strengths from $\sigma_{\ln(t)}=0.05$ to $0.2$. Energies
depending on disorder strength, are scanned from
$E=10^{-7}-10^{0}$. Here $N \gg \ell_{s}$. Inset figure shows the skewness of the data.
The number of samples is $10^4$ configurations.
\label{fig:SPS-D}}
 \efig
Energies are scanned from near the band center to the middle of
the conduction band, and far from the second anomaly in the band
edges. In fact, close to any anomaly (here a delocaliztion at the
band center), a violation of SPS will happen
[\onlinecite{Deych2}] in an interval close to the anomalous
state. It can be seen that for $\kappa \ll 1$, SPS is clearly
violated (non-SPS region). For $\kappa \gg 1$, independent of
disorder strength, standard SPS is restored again. In the SPS
spectral region, the localization length is a macroscopic length
as only one parameter of system. The conductance can be defined
as a function of this variable.

Both length scales $\ell_{s}$ and the localization length are
decreased when energy is swept from the vicinity of the band
center to the band edges. In energies close to zero, the new
length scale is greater than the localization length. There is a
critical energy where both length scales are of the same order
($\lambda\approx \ell_s$). Far from the band center, the
localization length is independent of energy and the new length
scale steeply decreases so that it would be much smaller than a
macroscopic localization length.

This result confirms the general conjecture of the authors
[\onlinecite{Deych}] that the second moment of the distribution
function of L.E. can be universally described in terms of
variables $\tau$ and $\kappa$ regardless of the microscopic nature
of the models under consideration. The form of the function
$\tau(\kappa)$ may differ for different models and its essential
behavior is not universal. All models follow $\tau(\kappa)=1 $
for $\kappa\gg 1$, while in the hopping disorder model, it is $\tau(\kappa)=2$.
In the model studied in the present paper, for
$\kappa\ll 1$, it can be seen an exceptional behavior compared to
other models such as Lloyd model
[\onlinecite{Deych,Deych2,Deych3}] and Anderson (onsite disorder)
and superlattice models [\onlinecite{Deych3}]. The scaling
parameter $\tau$ increases with $\kappa$ in the hopping disorder
model for $\kappa \ll 1$, while in the above models, $\tau$
steeply decreases with $\kappa$. As an example, analytical
calculations carried out in Ref.[\onlinecite{Deych}] for the
Lloyd Model produced $\tau=(\pi/2)\kappa$.

The power law form is the best fitted curve for $\kappa\ll1$.
 \beq
\frac{1}{\tau}\propto \alpha \kappa^{\beta} \eeq
 Coefficients are estimated by using a linear regression in log-log plot. The fitted power of
$\kappa$ and its coefficient are $\beta= 0.815\pm0.008$ and
$\alpha=1.64\pm0.03$, respectively. Since there is a kind of
delocalization at the band center, the L.E. sharply decreases near
the band center compared to $N\sigma_{\gamma}^{2}$
(Fig.(\ref{fig:Nvar-gam})). Therefore, the inverse scaling
parameter decreases in energies close to the band center.
%
%
 \bfig
\includegraphics[width=8 cm]{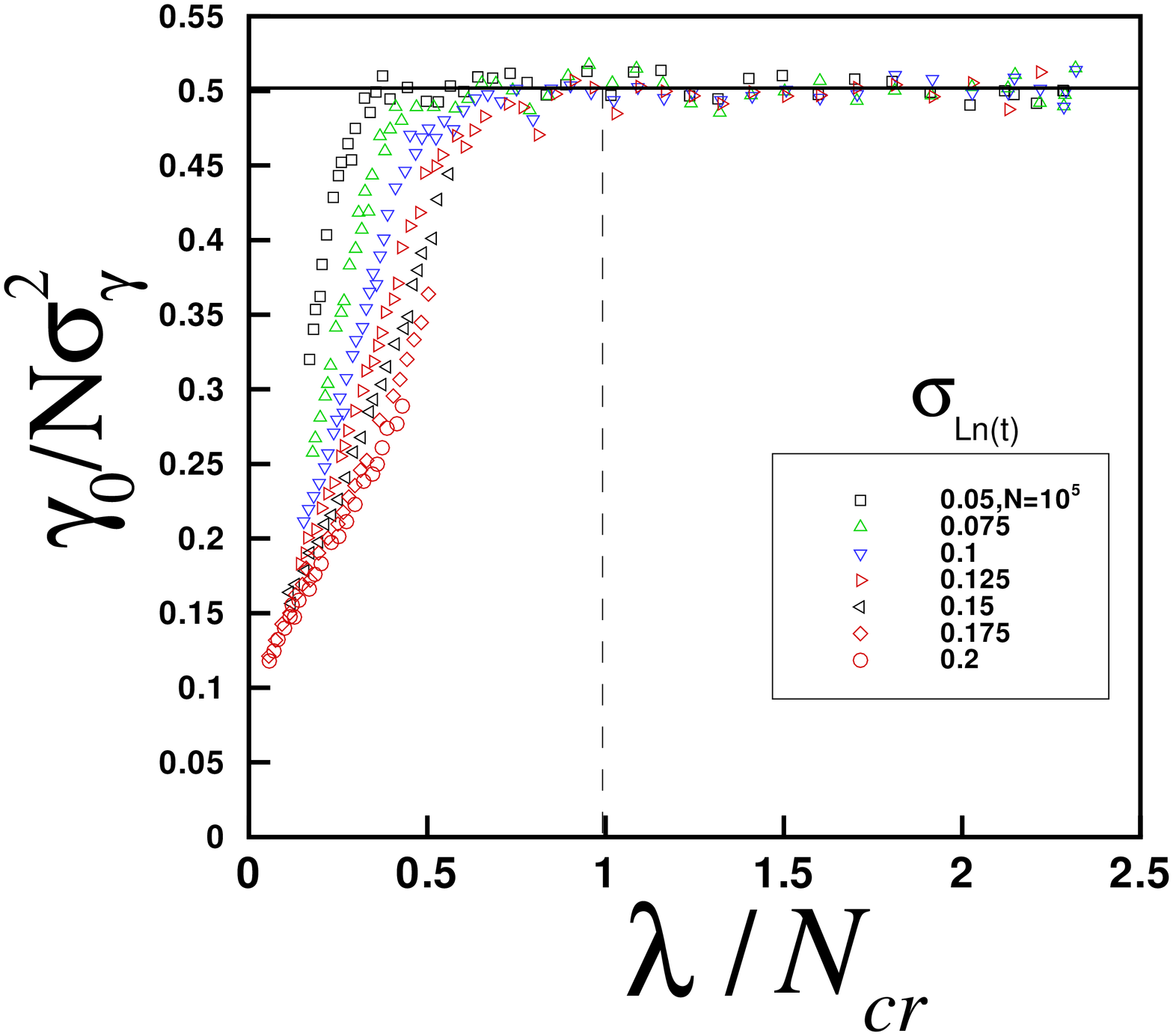}
\caption{ Inverse scaling parameter  $1/\tau =\gamma_{0}/(N
\sigma^{2}_{\gamma})$ versus localization length in the ratio of
the cross-over length ($\kappa' = \lambda/N_{cr}$) for different
disorder strengths. Energies
depending on disorder strength, are scanned from
$E=10^{-7}-10^{0}$.\label{fig:SPS}} \efig
\section{Cross-over Length as a physical interpretation of $\ell_{s}$}
Numerical results in Fig.(\ref{fig:Nvar-gam}) show that as a crude
approximation, one can consider the variance of L.E. independent
of energy in the non-SPS and SPS regimes (not near their
transition point). It was shown that the form of variance near the
band center is similar to its form in the band center
(Eq.(\ref{eq:unvar})). By using this form of the variance and the
scaling relations derived in
Eqs.(\ref{eq:cr-energy},\ref{eq:sps_scale}), the scaling parameter
function ($\tau$) can be proposed in the non-SPS and SPS regimes.

First, we investigate the scaling parameter in the non-SPS
regime. According to the infinite L.E. in the localized region
(Eq.(\ref{eq:cr-energy})) $\gamma_{0}\propto\sigma_{\ln(t)}^{2}
E^{\eta}$ and its variance  as
$N\sigma_{\gamma}^2\propto\sigma_{\ln(t)}^2$, it can be seen that
the measure of SPS (scaling parameter $\tau$) can be proposed to
have a linear relation with the dimensionless variable
$\kappa'=\lambda/N_{cr}$.

\beq\frac{1}{\tau} \propto E^{\eta} \propto
\frac{\lambda}{N_{cr}} \label{eq:proposed}\eeq

This equation shows a deviation from the SPS value (unity) in the
non-SPS regime. The cross-over length $N_{cr}$ plays the role of
a length scale like $\ell_{s}$ in this system. The above form for
the scaling parameter is independent of the system parameters
such as disorder strength when it is expressed in terms of $\tau$
and $\kappa'$. This expression is confirmed by Fig.(\ref{fig:SPS})
which shows the inverse scaling parameter versus $\kappa'$. It
can be seen that for $\kappa'\ll1$, all data for different
disorder strengths, coincide with each other on a single curve.
However, since in the second critical point ($\kappa'\approx1$),
the variance of L.E. is energy dependent and also, the scaling
form of L.E. (Eq.(\ref{eq:cr-energy})) is not correct in this
point (Fig.(\ref{fig:Nvar-gam})), curves with different disorder
strengths are separated from each other in the transition point.

In the SPS regime, the scaling of L.E. proposed in
Eq.(\ref{eq:sps_scale}) $\gamma_0\propto\sigma_{\ln(t)}^2$ and its
variance form as Eq.(\ref{eq:unvar}), show that the scaling
parameter $\tau$ is independent of system parameters and a
constant (Fig.(\ref{fig:SPS})).

Therefore, both the new length scale and the
cross-over length that are of the same order in the non-SPS regime,
can characterize the scaling properties. Fig.(\ref{fig:Ls})
compares these two length scales. It shows the ratio of the new
length to the cross-over length in terms of energy. In the Non-SPS
regime, they both weakly depend on energy, although, in the
transition region, the difference is remarkable.

As a result, the cross-over length is proposed as a physical and
meaningful interpretation for the new length scale ($\ell_{s}$).
On the other hand, we showed that the statistical distribution of
L.E. is different for sizes lower or greater than this length
scale.
%
%
\bfig
\includegraphics[width=8 cm]{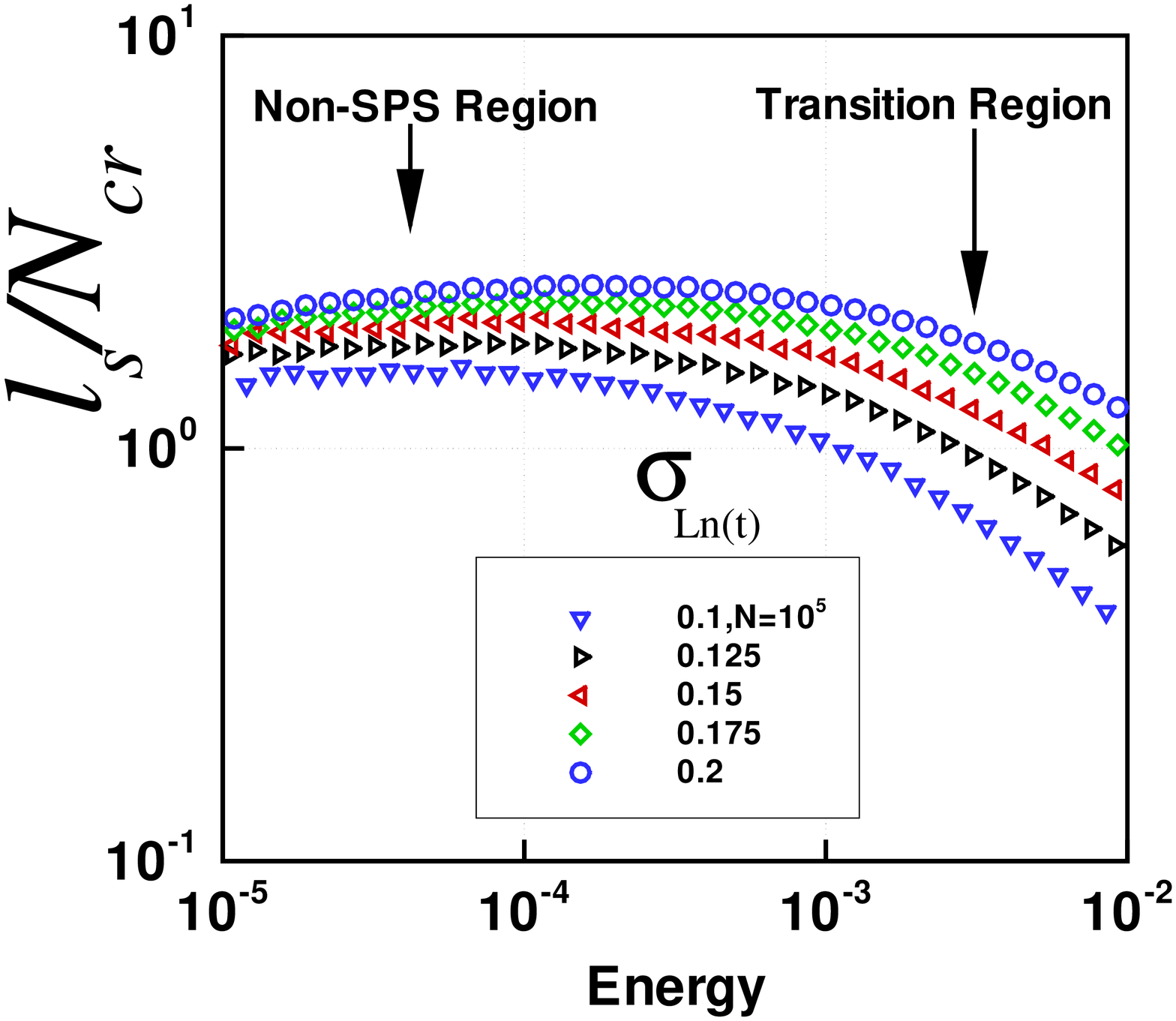}
\caption{The ratio of the new length scale ($\ell_{s}$) to the
cross-over length in terms of energy for different values of
disorder strength.\label{fig:Ls}}
 \efig
\section{Conclusion}
In this paper, we study the scaling theory in the hopping
disorder model. The main result of this paper is to show the
single parameter scaling (SPS) is violated not only in a region
with chiral symmetry, but also in the localized region where there
exists a standard symmetry class. The localized region is also
divided into two regimes: SPS and non-SPS regimes. We proposed
the scaling relations for the Lypunov Exponent in these two
regimes. The criterion of the SPS is controlled by a new length
scale which is related to the integral of density of states,
$\ell_{s}$ defined in Ref.[\onlinecite{Deych}]. The SPS holds when
the localization length $\lambda$ exceeds the new length
($\lambda\gg\ell_{s}$). In $\lambda\ll\ell_{s}$ regime, standard
deviation of the Lyapunov Exponent ( $\gamma$) distribution can
be described by two independent scaling parameters: the mean of
$\gamma$ and $\kappa=\lambda/\ell_{s}$.

We showed that all data related to the variance and mean of the
Lyapunov Exponent with different values of disorder strengths,
system sizes and also the data extracted from various energy
regions, lie on a single curve, when they are expressed in terms
of the inverse scaling parameter
$1/\tau=\gamma_{0}/N\sigma^2_{\gamma}$ and the dimensionless
variable $\kappa$.

The cross-over length ($N_{cr}$) which separates the region with
chiral symmetry from that of standard symmetry, is proposed as a
meaningful physical interpretation for $\ell_{s}$.

\begin{acknowledgments}
I wish to acknowledge Prof. Keivan Esfarjani for a critical reading of the manuscript
and his effective comments and suggestions.
I would like to thank Dr. S. Mahdi Fazeli for useful discussions on the initial
manuscript. I would like also to thank Dr. Nima Ghal-Eh for his
review and editing of the manuscript.
\end{acknowledgments}

\end{document}